# Análise das estratégias de planejamento de tempos de voo pelas companhias aéreas

Ana Beatriz Rebouças Eufrásio
Alessandro V. M. Oliveira↛
Instituto Tecnológico de Aeronáutica, São José dos Campos, Brasil
↛ Autor correspondente. Instituto Tecnológico de Aeronáutica. Praça Marechal Eduardo Gomes, 50. 12.280-250 - São José dos Campos, SP - Brasil.
E-mail: alessandro@ita.br.

*Resumo*: Este estudo explora as abordagens utilizadas pelas companhias aéreas na definição dos tempos de voo. Ressalta-se a necessidade de equilibrar fatores operacionais e estratégicos, como a otimização do uso de recursos – incluindo aeronaves, tripulação e combustível – e o gerenciamento dos riscos relacionados a atrasos e congestionamentos. O trabalho detalha uma análise nacional focada em voos domésticos, investigando os fatores que influenciam as empresas a ajustar os tempos de voo programados e o impacto dessa prática na pontualidade. Os resultados indicam que decisões sobre o tempo de voo são influenciadas tanto por aspectos operacionais quanto estratégicos, sendo afetadas pela concorrência e pelas políticas de pontualidade e alocação de slots do setor. Ademais, constatou-se que a adição de tempo extra é uma estratégia eficaz para reduzir atrasos, embora possa ocultar deficiências do sistema.

*Palavras-chave*: transporte aéreo, companhias aéreas, atrasos de voo.

## I. Introdução

O planejamento da programação de voos é uma das principais atividades das companhias aéreas, estando diretamente relacionado à gestão dos custos operacionais e da pontualidade. Por um lado, tempos de voos mais longos reduzem os riscos associados a atrasos e congestionamentos. Por outro, reduzir o tempo de voo possibilita otimizar o uso de certos recursos, como aeronaves e tripulação. O combustível também exerce um importante papel nessa relação: preços mais altos de combustível podem incentivar as companhias a buscar reduzir seu consumo, por exemplo adotando menores velocidades de cruzeiro e consequentemente aumentando os tempos de voo.

Nos últimos anos tem-se observado que, apesar da avançada tecnologia das aeronaves, as durações dos voos em algumas rotas aumentaram (Fan, 2019). Segundo uma matéria publicada no portal Which em 2018, voos do aeroporto de Gatwick em Londres para o aeroporto JFK em Nova Iorque, por exemplo, duraram em média 20 minutos a mais em 2018, em comparação aos seus tempos de voo programados de 2008. Já os voos de Heathrow para Newark demoraram cerca de 35 minutos a mais, comparando também 2018 com 2008. Paralelamente, e talvez não coincidentemente, as companhias aéreas atingiram altos níveis de pontualidade. De acordo com uma reportagem do portal USA Today de 2013, nos Estados Unidos, em 2012, pela primeira vez ocorreram mais chegadas antecipadas do que atrasos/cancelamentos de voos.

Programar tempos de voos maiores de forma a proporcionar à companhia aérea mais flexibilidade para lidar com possíveis atrasos e ainda cumprir o horário de chegada planejado é uma prática conhecida no setor como schedule padding (Yimga & Gorjidooz, 2019). Nos últimos anos, as companhias aéreas vêm sendo acusadas de adotar agressivamente essa prática como forma de melhorar seus índices de pontualidade. Esse tempo extra (aqui denominado de buffer estratégico) pode estar ocultando ineficiências do sistema, como espaço aéreo mais congestionado, taxiamentos mais longos e até maior consumo de combustível. Além disso, esse comportamento manipulativo pode prejudicar a percepção dos passageiros sobre a qualidade do serviço das companhias aéreas (Yimga & Gorjidooz, 2019).

No entanto, tempos de voo mais longos podem ser uma consequência inevitável de ajustes operacionais e não apenas uma estratégia para evitar atrasos. Um dos principais parâmetros para determinar a velocidade de cruzeiro de um voo é o Cost Index (CI), que representa a relação entre custos dependentes do tempo (como tripulação e aeronave) e o custo com combustível. Quando o custo unitário dos insumos dependentes do tempo é baixo com relação ao custo com combustível, as companhias aéreas tendem a priorizar durações de voos maiores, adotando uma velocidade menor de cruzeiro e consequentemente consumindo menos combustível. Já se os custos dependentes do tempo forem maiores que o custos com combustível, os voos tendem a ter durações menores e um maior consumo de combustível (Young, 2018; Deo, Silvestre & Morales, 2020). Assim, nem todo tempo extra adicionado é um buffer estratégico.

A busca pelo equilíbrio entre esses fatores é um dos grandes desafios do planejamento da programação de voos. Este trabalho busca dissociar os determinantes estratégicos e operacionais dos tempos extras adicionados aos tempos de voo – assim, dissociando também, fatores operacionais da prática do schedule padding. Analisa-se como o tema é explorado na literatura internacional e também em um estudo nacional, que investiga o caso das companhias aéreas brasileiras de 2001 a 2018. Além de analisar os determinantes do tempo extra, esse estudo também analisa a eficiência do aumento dos tempos de voo como estratégia para melhorar a pontualidade, estimando seu impacto na probabilidade de ocorrência de atrasos para o caso brasileiro.

No Brasil, o setor aéreo presenciou nos últimos anos grandes variações de índices de pontualidade. Enquanto a ocorrência de atrasos no mercado brasileiro reduziu de 27,5% em 2008 para 15,8% em 2018, os casos de voos com chegada antecipada aumentaram de 0,4% para 44,3% no mesmo período (segundo dados do Voo Regular Ativo, ANAC), como apresentado na Figura 1. Além disso, o aeroporto de Guarulhos em São Paulo (GRU), um dos principais aeroportos internacionais do país, foi considerado um dos 10 aeroportos mais atrasados do mundo em 2008. Já em 2018, o mesmo aeroporto foi classificado em 10° lugar em um ranking que apresentava os 20 aeroportos com melhores índices de pontualidade do mundo, de acordo com sua categoria . Paralelamente, observa-se que, ao longo dos anos, os tempos de voo programados de voo de algumas rotas domésticas aumentaram enquanto os tempos de voo reais dos voos

diminuíram – ou cresceram menos que os tempo de voos programados, como mostra a (Figura 2).

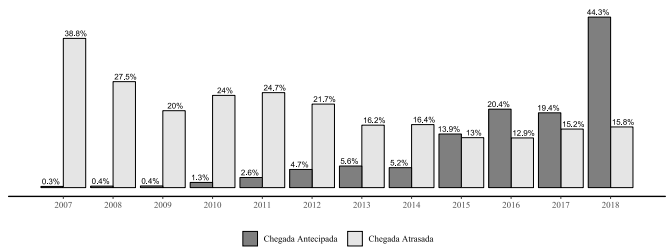

**Figura 1 - Evolução da proporção de chegadas antecipadas e de chegadas atrasadas no Brasil (Dados do VRA ANAC com cálculos próprios)**

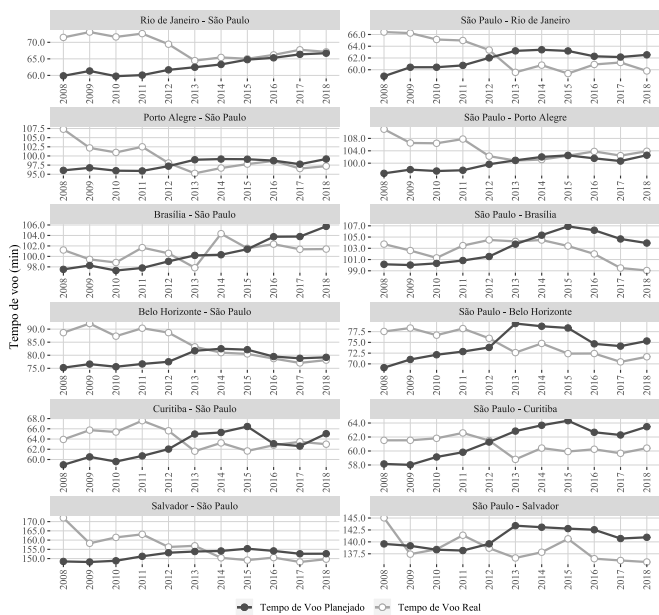

**Figura 2 – Evolução do tempo real médio de voo e do tempo programado médio de voo para as 12 rotas domésticas mais densas de 2018 rotas selecionadas (Dados do VRA ANAC com cálculos próprios)**

Essas características levantam algumas questões, exploradas neste trabalho: Por quais motivos os tempos de voos estão aumentando em algumas rotas? As companhias aéreas aumentam seus tempos de voo como estratégia para evitar atrasos? Ou seriam esses aumentos um ajuste às condições operacionais? Além disso, a competição tem algum impacto na decisão do tempo de voo programado? E como as políticas de pontualidade e de slots influenciam no tempo extra a ser adicionado e nos índices de pontualidade? O aumento de tempo de voo traz efetivamente uma redução de atrasos?

Esse trabalho mostra a importância de se identificar situações que favoreçam o aumento de um buffer estratégico e as situações que tornam necessária a adição de um tempo extra como ajuste operacional. Compreender o que motiva o aumento do tempo de voo pode ser extremamente útil para o agente regulador. Isso permite que seja investigado se os índices de pontualidade refletem de fato o desempenho de cada companhia na gestão de seus tempos de voo ou se existe uma melhoria induzida por um aumento do tempo de voo não-operacional.

A análise sugerida tem como objetivo demonstrar a importância da diferenciação entre aumentos de tempo de voo estratégicos e operacionais. Inicialmente, serão apresentados os principais conceitos associados à programação de voos e à determinação dos tempos de voo. Em seguida, será discutido como o tema tem sido abordado na literatura internacional e a apresentação de um estudo nacional.

## II. O PLANEJAMENTO DA PROGRAMAÇÃO E DOS TEMPOS DE VOO

O planejamento da programação de voos de uma companhia aérea envolve a determinação das frequências, durações de voos, horários de partidas, rotas e conexões, além da atribuição das tripulações e aeronaves. É um processo interativo e complexo, e as companhias precisam organizar os recursos que têm disponíveis para atender à demanda de toda sua malha (Holloway, 2008). Algumas características tornam esse planejamento ainda mais difícil, como o clima e congestionamentos – externalidades que nem sempre podem ser previstas ou evitadas.

Uma etapa do planejamento da programação é a definição do tempo bloco-a-bloco programado, que representa o tempo entre a partida programada do portão do aeroporto de origem até a chegada programada ao portão do aeroporto de destino. É importante ressaltar que, além do tempo bloco-a-bloco, a companhia também precisa definir o seu tempo de turnaround, que se refere ao tempo dedicado às operações de solo, como a manutenção da aeronave e operações de embarque e desembarque. A definição do tempo bloco-a-bloco e do tempo de turnaround estão diretamente relacionadas ao tipo de aeronave atribuída, além da definição do sequenciamento de rotas para cada aeronave da frota.

Como apresentado na Figura 3, na prática, os horários de chegada e de partida não ocorrem necessariamente como o planejado. O tempo bloco-a-bloco real pode ser diferente do programado, com a ocorrência de atrasos na partida ou na chegada e ainda possíveis variações no tempo da aeronave no ar.

**Tabela 2 – Determinantes dos tempos de voo**

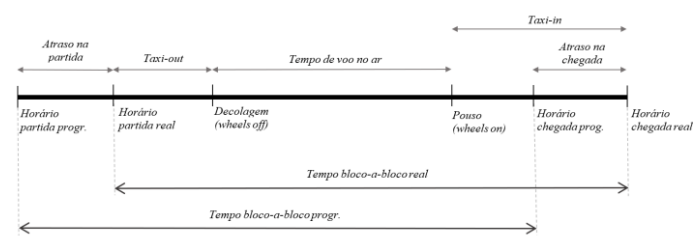

**Figura 3 - Decomposição do tempo bloco-a-bloco de um voo (adaptado de Deshpande & Arıkan, 2012)**

O cálculo do tempo bloco-a-bloco programado é apresentado na equação (1) e do tempo bloco-a-bloco real na equação (2):

```
Tempo bloco-a-bloco programado = Horário chegada
   programada – Horário partida programada  (1)

Tempo bloco-a-bloco real = Horário chegada real –
   Horário partida real (2)
```

Pode-se assumir que existe um tempo bloco-a-bloco "desimpedido", denominado também de tempo bloco-a-bloco "base". Esse tempo representa o tempo bloco-a-bloco da viagem em condições ideais, livre de qualquer externalidade. Considera-se que o tempo bloco-a-bloco desimpedido pode ser representado por um dos menores tempo bloco-a-bloco registrados, como realizado por Fan, (2019), Yimga e& Gorjidooz (2019), Kafle e& Zou (2016), Zou e& Hansen (2012) e Mayer e Sinai (2003a) – esses estudos consideram um percentil baixo da distribuição dos tempos bloco-a-bloco reais para representar o tempo bloco-a-bloco desimpedido, como o 5º, 10º ou 20º percentil para cada combinação de rota, companhia aérea e aeronave em um certo período de tempo. Zou & Hansen (2012) aconselham não utilizar o mínimo absoluto dos tempos bloco-a-bloco reais para tornar o cálculo dos tempos extras robustos à possíveis erros de registro e também à influência de condições incomuns e extremamente favoráveis ao voo – por exemplo, situações excepcionais com fortes ventos de cauda, não usuais na rota, que aumentam a velocidade da aeronave e podem ocasionar em durações de voo menores.

O tempo bloco-a-bloco programado é, então, a soma de um tempo extra programado, adicionado por segurança para lidar com os riscos associados ao não cumprimento da programação. Dessa forma, o tempo extra programado é definido por (3):

```
Tempo extra programado = Tempo bloco-a-bloco programado
 - Tempo bloco-a-bloco desimpedido (3)
```

Diversos fatores – tanto internos como externos à companhia aérea – justificam a variação do tempo bloco-a-bloco real. Pode-se citar condições atmosféricas adversas, atrasos e cancelamentos de voos, congestionamentos e atrasos nas operações de embarque e desembarque. A seguir é apresentada uUma discussão detalhada sobre como a literatura internacional tem analisado o impacto de fatores operacionais e estratégicos nos tempos de voo – e o efeito disso no desempenho das companhias aéreas.

### III. Revisão da Literatura

Muitos pesquisadores buscam analisar quais são os efeitos das variações de tempo de voo na pontualidade, um dos principais indicadores da qualidade de serviço de uma companhia aérea, e também nos custos operacionais. No entanto, alguns aspectos dessas relações são controversos e há pouca discussão sobre a dissociação de motivações operacionais e estratégicas nos aumentos dos tempos bloco-a-bloco programados.

Maiores tempos de voos programados auxiliam a companhia aérea a acomodar possíveis eventos inesperados, o que pode gerar um impacto positivo na qualidade de serviço. Voos mais longos provavelmente geram melhores índices de pontualidade, atraindo mais passageiros devido à garantia de menos atrasos (Kang e Hansen, 2017; Prince e Simon, 2009; Skaltsas, 2011). Porém, do ponto de vista do passageiro, durações menores de voos podem ser consideradas uma vantagem competitiva, especialmente em rotas com competição intensa (Skaltsas, 2011). Yimga e Gorjidooz (2019) associaram a prática do schedule padding a efeitos negativos sobre o bem-estar do consumidor. Forbes, Lederman e Yuan (2018) sugerem que mesmo essa prática trazendo mais confiabilidade ao horário de chegada do voo, o tempo total da viagem aumenta, não sendo diretamente evidente se a qualidade do serviço também aumenta.

Variar o tempo de voo pode ser também uma estratégia para lidar com a competição. Kang e Hansen (2017) observam que em rotas competitivas, as companhias aéreas tendem a aumentar seus tempos bloco-a-bloco programados. Prince e Simon (2009) e Fan (2019) observam que as companhias tendem a reduzir seus tempos de viagem programados em mercados menos competitivos, tornando-se mais vulneráveis a atrasos, nesse caso. Considerando a competição entre qualidade e preço, Forbes, Lederman e Wither (2019) observam que quando uma companhia é incentivada a melhorar seus índices de pontualidade, seus competidores podem ser induzidos a adotar a mesma estratégia.

Apesar do aumento na duração da viagem tornar a operação mais ajustável e aumentar a probabilidade de o voo ser pontual, tempos de voos maiores podem representar uma subutilização dos recursos da companhia. Mayer e Sinai (2003b) e Forbes, Lederman e Yuan (2018) sugerem que um tempo bloco-a-bloco programado maior está associado a maiores custos com a tripulação e a um uso menos eficiente da aeronave, ao atribuir menos missões para a frota. Para Fan (2019), quando não há pressões para ser pontual, a companhia aérea busca reduzir seus tempos de voo. Já para Prince e Simon (2009), maiores tempos de voo programados podem reduzir os custos por reduzir a necessidade de se melhorar a pontualidade, por exemplo com aeronaves mais rápidas e com a contratação de mais funcionários. Miranda e Oliveira (2018) observam que a adição de tempos extras está relacionada com a reduções da ocorrência de atrasos e cancelamentos, mas não observaram evidências de que essa prática aumenta os custos operacionais que poderiam ser repassados aos consumidores através da tarifa da passagem. Para Fan (2019), a decisão de adicionar um tempo extra deve considerar o trade-off que existe entre usar mais recursos para aumentar a flexibilidade da operação com maiores tempos de voo, ou usar mais recursos quando ocorrer um atraso.

Além disso, a previsibilidade do tempo de voo é um dos principais fatores a serem considerados no planejamento da duração do voo. Quanto menos previsível, ou seja, quanto mais incerto for o tempo bloco-a-bloco real, maior o tempo bloco-a-bloco programado tende a ser (Kang e Hansen, 2017; Brueckner, Czerny e Gaggero, 2019). Por exemplo, rotas operadas em regiões com condições atmosféricas adversas podem ter o seu tempo bloco-a-bloco real mais incerto. As companhias aéreas variam seus tempos bloco-a-bloco programados de acordo com a sazonalidade, direções dos ventos e até com a horário do dia, considerando possíveis congestionamentos (Holloway, 2008). Hao e Hansen (2014) mostram que baixos índices de pontualidade em um ano geralmente aumentam o tempo bloco-a-bloco programado do ano seguinte.

Além disso, uma companhia aérea pode decidir programar um tempo bloco-a-bloco maior (ou seja, adicionar mais tempo extra ao seu tempo de voo programado) como forma de lidar com restrições da infraestrutura. Para lidar com os riscos associados a congestionamentos no aeroporto e no espaço aéreo, as companhias podem optar por programar um tempo bloco-a-bloco maior, para garantir as conexões de passageiros e evitar atrasos e cancelamentos de voos (Fan, 2019). Há também o efeito cascata do atraso: um pequeno atraso em um voo que ocorreu mais cedo pode ser propagado para outros voos, prejudicando seus índices de pontualidade (Kafle & Zou, 2016).

O tempo de voo também impacta o consumo de combustível e a eficiência energética do voo. O custo com combustível é um dos principais componentes dos custos operacionais de uma companhia aérea e sua importância aumenta na medida que os preços com combustível aumentam (Şafak, Atamtürk, &e Aktürk, 2019). Custos maiores com combustível incentivam companhias aéreas a melhorarem sua eficiência energética (Zou et al., 2014). Assim, custos de insumos que dependem do tempo de voo, como tripulação e manutenção, aumentam com maiores tempos de viagem, enquanto o consumo de combustível se reduz com melhores velocidades (Edwards, Dixon-Hardy &e Wadud, 2016). O Cost Index (CI) – a relação entre custos variáveis com o tempo e custo com combustível por voo – permite determinar a velocidade ótima de cruzeiro, correspondente aos menores custos operacionais (Young, 2018). Menores CI apontam para menores velocidades de cruzeiro e assim menos consumo de combustível. Fan (2019) observa que quando o preço de combustível aumenta, as companhias podem decidir adotar menores velocidades de cruzeiro como forma de reduzir o consumo de combustível. No entanto, o autor notou um impacto pouco relevante do preço de combustível nos tempos bloco-a-bloco.

Além dos fatores estratégicos e operacionais, políticas e regulações do setor aéreo podem influenciar as decisões das companhias aéreas com relação às durações das viagens. Nos Estados Unidos, a regulação de divulgação dos índices de pontualidade vem sendo implementada desde 1987 pelo Departamento de Transportes. Com isso, os índices de pontualidade das principais companhias aéreas do país passaram a ser públicos. Shumsky (1993) encontram evidências de que os tempos de voo programados de algumas rotas domésticas dos Estados Unidos aumentaram como resposta à essa regulação. Forbes, Lederman e Wither (2019) observam que ao ter que reportar seus índices de pontualidade, as companhias tendem a aumentar seus tempos de voo programados. Em alguns casos, os tempos de voo reais não se alteraram, enquanto os tempos programados aumentaram. A adoção de slots nos aeroportos também influencia o gerenciamento da pontualidade das companhias aéreas. Santos e Robin (2010) observam que os atrasos são maiores em aeroportos coordenados e menores em aeroportos eslotados.

Mudanças e inovações na Gestão de Tráfego Aéreo também podem trazer ajustes nas durações das viagens. Programas como o Next Generation Air Transportation System (NextGen) nos Estados Unidos e o Single European Sky ATM Research (SESAR) na Europa correspondem à implementação de um sistema de navegação mais eficiente. Esses programas não se baseiam unicamente em auxílios de solo, com um sistema de navegação por satélite, e adotam também a Navegação com Base em Performance (PBN – Performance Based Navigation) para obter informações de posições das aeronaves mais precisas. Diana (2017) aponta que o programa NextGen e a reestruturação do espaço aéreo dos Estados Unidos aumentaram os índices de pontualidade das companhias aéreas.

## IV. O Caso Brasileiro

No Brasil, Eufrásio, Eller & Oliveira (2021) investigaram as motivações e os efeitos da adoção de tempos extras nos tempos bloco-a-bloco programados. Para isso, dois modelos econométricos foram construídos: um para estimar os efeitos de fatores estratégicos e operacionais na quantidade de tempo extra adicionado e outro para analisar impacto desse tempo extra nos índices de pontualidade. Considerou-se na amostra os voos domésticos realizados por companhias aéreas brasileiras entre 2001 e 2018 para o transporte de passageiros e entre capitais. Os dados utilizados correspondem a 322 rotas, agregados por mês e por pares de cidade. Os dados operacionais das companhias aéreas foram obtidos nas bases de dados da Agência Nacional de Aviação Civil (ANAC), disponíveis publicamente on-line. Os dados de preço de combustível são foram coletados nos dados disponibilizados on-line pela Agência Nacional de Petróleo, Gás Natural e Biocombustível (ANP).

No primeiro modelo do estudo, o objetivo é investigar o efeito de fatores operacionais e estratégicos no tempo extra de voo. Para cada rota em cada mês, determinou-se um tempo bloco-a-bloco desimpedido. O tempo bloco-a-bloco desimpedido considerado foi o 5º percentil da distribuição dos tempos bloco-a-bloco reais para cada rota/companhia aérea/aeronave/mês. Em uma análise de robustez, testou-se também o modelo com o 10º percentil – não havendo mudanças significativas nos resultados. Com isso, calculou-se o tempo bloco-a-bloco extra médio (de acordo com a equação 3), subtraindo os tempos bloco-a-bloco programados médios do tempo bloco-a-bloco desimpedido médio de cada rota.

No primeiro modelo do estudo, o objetivo é investigar o efeito de fatores operacionais e estratégicos no tempo extra de voo. O modelo econométrico estimou o impacto de cada fator no tempo bloco-a-bloco extra, considerando seus valores defasados em três meses. Essa defasagem foi adotada por se considerar que o tempo bloco-a-bloco extra faz parte do planejamento do tempo de voo, e é decidido com antecedência, com base nos resultados operacionais passados. Também se testou o modelo com uma defasagem de 12 meses, não havendo nenhuma mudança significativa nos valores estimados. Foram adicionados controles de direção/sazonalidade/localidade para controlar na estimação o potencial efeito da direção e intensidade dos ventos nos tempos de voo.

Analisando inicialmente os ajustes realizados no tempo bloco-a-bloco programado devido a fatores operacionais, o modelo sugere que as companhias aéreas tendem a adicionar mais tempo extra quando o custo de combustível aumenta. Esse resultado indica que um aumento no preço de combustível provoca uma queda no Cost Index esperado para a operação futura, levando as companhias a reduzirem suas velocidades de cruzeiro, aumentando assim o tempo de voo e reduzindo o consumo energético – possivelmente buscando reduzir custos operacionais, de acordo com a mudança do Cost Index. Outro fator operacional relevante é a frequência de voos na rota. Os resultados apontam que rotas com maior frequência de voos reduzem o tempo extra adicionado aos tempos bloco-a-bloco, sugerindo que esses voos apresentam maior eficiência operacional, com menores custos variáveis no tempo com relação aos custos de combustível. Além disso, os resultados apontam que as companhias aéreas tendem a adicionar mais tempo extra às rotas operadas em aeroportos que tiveram, no período defasado analisado, maiores índices de atrasos – indicativo escolhido para representar o efeito cascata dos atrasos, ou seja, o impacto que um atraso em um voo pode causar em outro voo.

Com relação aos fatores estratégicos, os resultados sugerem que as companhias programam tempos bloco-a-bloco maiores, em rotas mais competitivas, possivelmente como forma de lidar com a maior competição da qualidade do serviço. Além disso, quanto maior a presença de companhias aéreas low-cost (LCCs), mais as companhias que operam a rota tendem a aumentar seus tempos bloco-a-bloco extras. Esses resultados indicam a ocorrência do schedule padding, ou seja, de um aumento do tempo bloco-a-bloco devido a fatores estratégico, não atribuíveis a ajustes operacionais – tempos extras que são, portanto, buffers estratégicos.

Há fatores que podem ser considerados tanto operacionais como estratégicos, como a densidade da rota (número de passageiros transportados). A densidade é considerada um fator estratégico já que as companhias podem tender a se preocupar mais com a satisfação de seus passageiros em rotas mais densas – nesses casos, os atrasos atingem um público maior, podendo trazer um maior prejuízo para a sua reputação. Além disso, a densidade pode também ser considerada um fator operacional já que rotas mais densas tendem a ter mais passageiros em conexão, tornando a operação mais complexa e possivelmente levando as companhias aéreas a aumentarem seus tempos bloco-a-bloco. Outro fator considerado estratégico e operacional é a incerteza do tempo de voo, representado pelo coeficiente de variação da distribuição dos tempos bloco-a-bloco de uma rota em um certo tempo. Observa-se que mais tempo extra tende a ser adicionado em voos cujos tempos bloco-a-bloco são mais incertos.

O segundo modelo do estudo de Eufrásio, Eller & Oliveira (2021) tem como objetivo principal analisar o impacto do tempo extra nos índices de atraso. Observa-se que a adição de um minuto extra no tempo bloco-a-bloco reduz em aproximadamente 13% a probabilidade de atraso em uma rota.

No entanto, essa redução tende a ser menos efetiva nos aeroportos com regimes de slots – possivelmente as companhias aéreas têm menos flexibilidade na programação de voos que partem ou pousam nesses aeroportos. O modelo também observa que rotas mais densas e rotas operadas em aeroportos com pistas mais congestionadas tem maior probabilidade de atrasarem. Como o modelo 1 sugere que a densidade aumenta o tempo extra adicionado, esse resultado aponta que, mesmo com o aumento dos tempos bloco-a-bloco, rotas mais densas ainda assim atrasam mais.

Já a competição tem seu efeito na probabilidade de atraso dissipado com a inclusão da variável do tempo extra no modelo, enquanto rotas com maior participação de LCCs tendem a ter menos atrasos. No modelo 1 havia sido observado que rotas com maior presença de LCCs tendem a adicionar mais tempo extra – assim, em situações em que as companhias decidem aumentar seus tempos bloco-a-bloco devido a uma maior competição com LCCs (possivelmente buscando cortes de custos), também são observadas melhorias na pontualidade da rota.

São analisados também o impacto de diferentes reformas regulatórias nos tempos extras: resolução sobre a divulgação dos percentuais de atrasos em 2012 , nova regulação sobre o regime de slots nos aeroportos brasileiros em 2014 e a implementação do PBN em alguns aeroportos brasileiros, iniciada em 2010. Com relação às regras de divulgação das estatísticas de atrasos no Brasil, antes de sua adoção, em 2012, as companhias tinham uma tendência de reduzir os tempos extras adicionados – possivelmente as companhias toleravam mais menores índices de pontualidade. Após a introdução dessa regulação, a tendência se tornou o oposto: as companhias passaram a adicionar mais tempos extras. Observa-se, portanto, que essa regulação intensificou a prática de schedule padding, já que é independente de mudanças operacionais e está associado ao posicionamento estratégico das companhias com relação à divulgação das estatísticas de pontualidade.

Com relação aos slots, os resultados do estudo sugerem que as rotas que partem ou chegam em aeroportos slotados tem um menor tempo extra adicionado e uma maior probabilidade de atrasar. No entanto, com a implementação de uma regulação de slots mais rigorosa em 2014, houve uma redução no tempo extra adicionado e também na probabilidade de atraso – esse resultado aponta que a regulação foi efetiva na redução de atrasos, sem incentivar a adição de tempos extras aos tempos bloco-a-bloco. Já a adoção das operações PBN em alguns aeroportos brasileiros a partir de 2010 não foi considerada estatisticamente significante para a determinação dos tempos extras nem para a probabilidade dos atrasos. Uma possível justificativa – não explorada pelo estudo apresentado – é que o PBN pode ter impactado não apenas o tempo bloco-a-bloco programado, mas também o tempo bloco-a-bloco desimpedido, sem ter gerado grandes impactos no tempo extra.

## V. Conclusões

Portanto, pode-se dizer que as companhias aéreas aumentam seus tempos de voo para atrasar menos? Os resultados do estudo de Eufrásio, Eller & Oliveira (2021) para o caso brasileiro sugerem que parte do aumento do tempo bloco-a-bloco acontece por ajustes operacionais, como variações no "cost index" das aeronaves, mas também por motivações estratégicas. Foram encontradas evidências de que o tempo extra adicionado reduz a probabilidade de atrasos nas rotas – a estimativa é que a adição de 1 minuto ao tempo bloco-a-bloco programado reduz em 13% a probabilidade de atraso na rota.

Na decomposição do tempo bloco-a-bloco em fatores operacionais e estratégicos, a significância dos fatores estratégicos indica a adoção da prática de schedule padding pelas companhias aéreas – ou seja, aumentos nos tempos de voo programados com a adição de buffers estratégicos como forma de adquirir vantagens competitivas com a redução de atrasos. Dentre os fatores estratégicos, os resultados sugerem que uma maior competição e uma maior presença de LCCs na rota incentivam a adição de tempos extras. A regulação sobre a publicação dos índices de pontualidade, válida desde 2012, também tem motivado as companhias aéreas a adotarem esse comportamento. Após a implementação dessa regulação, as rotas apresentam não apenas maiores tempos extras, mas também uma menor probabilidade de atraso. Já o novo regime de slots implementado em 2014 aparentemente inibiu o aumento nos tempos bloco-a-bloco programados. O modelo também identificou que as companhias aumentam as durações de voo programadas em rotas mais densas – porém, ainda assim, rotas mais densas têm maior probabilidade de atraso.

Uma das principais contribuições de Eufrásio, Eller & Oliveira (2021) foi a decomposição do tempo bloco-a-bloco extra em fatores operacionais e estratégicos, auxiliando a identificar características que incentivam o schedule padding. O problema dessa prática é que a adição de buffers estratégicos enfraquece a busca por uma maior eficiência operacional, mascarando ineficiências do sistema, como congestionamentos e a propagação de atrasos, e induzindo um uso menos eficaz dos recursos das companhias aéreas. Dessa forma, as estatísticas de pontualidade não são necessariamente confiáveis – podem não estar refletindo efetivamente o desempenho das companhias na gestão dos tempos de voo e de seus recursos.